\documentclass[12pt]{article}
\usepackage{graphicx}

\newcommand\pubnumber{SNSN-323-63}
\newcommand\pubdate{\today}

\def\ific{Instituto de F\'{\i}sica Corpuscular (IFIC)$,$
	CSIC-Universitat de Val\`encia$,$ \\  
	Apartado de Correos 22085$,$ E-46071 Valencia$,$ Spain}

\def\support{\footnote{Work supported by the European Union grants H2020-MSCA-ITN-2015/674896-Elusives and H2020-MSCA-RISE-2015/690575-InvisiblesPlus, by the Spanish MINECO through grants 
FPA2014-57816-P and  SEV-2014-0398,
and by Generalitat Valenciana grant PROMETEO/2014/050.}}

\def\Title#1{\begin{center} {\Large #1 } \end{center}}
\def\Author#1{\begin{center}{ \sc #1} \end{center}}
\def\Address#1{\begin{center}{ \it #1} \end{center}}

\newcommand\pubblock{\rightline{\begin{tabular}{l} \pubnumber\\
         \pubdate  \end{tabular}}}
\newenvironment{Abstract}{\begin{quotation}  }{\end{quotation}}
\newenvironment{Presented}{\begin{quotation} \begin{center} 
             PRESENTED AT\end{center}\bigskip 
      \begin{center}\begin{large}}{\end{large}\end{center} \end{quotation}}
\def\Acknowledgements{\bigskip  \bigskip \begin{center} \begin{large}
             \bf ACKNOWLEDGEMENTS \end{large}\end{center}}

\def\beq{\begin{equation}}
\def\eeq#1{\label{#1}\end{equation}}
\def\eeqn{\end{equation}}

\def\beqa{\begin{eqnarray}}
\def\eeqa#1{\label{#1}\end{eqnarray}}
\def\eeqan{\end{eqnarray}}

\let\bar=\overbar

\def\Dslash{\not{\hbox{\kern-4pt $D$}}}
\def\dslash{\not{\hbox{\kern-2pt $\del$}}}

\def\msb{{\bar{\ssstyle M \kern -1pt S}}}

\begin{document}
\begin{titlepage}
\pubblock

\vfill
\Title{Atmospheric neutrinos and new physics}
\vfill
\Author{Nuria Rius \support}
\Address{\ific}
\vfill
\begin{Abstract}
 We discuss recent searches for new physics using  high-energy atmospheric neutrino data from  IceCube,
 namely sterile neutrinos with masses in the  range $\Delta m^2 = 0.01$ eV$^2$ - 10 eV$^2$, and 
 non-standard interactions (NSI) in the $\nu_\mu - \nu_\tau$ sector. We also present a 
  brief review of the current status of NSI theory and phenomenology.
\end{Abstract}
\vfill
\begin{Presented}
NuPhys2016, Prospects in Neutrino Physics\\
Barbican Centre, London, UK,  December 12--14, 2016
\end{Presented}
\vfill
\end{titlepage}
\def\thefootnote{\fnsymbol{footnote}}
\setcounter{footnote}{0}

\section{Introduction}

Neutrino oscillation experiments have established that neutrinos are massive, which in turn requires physics 
beyond the Standard Model (SM). The new physics scale is unknown, but if 
sufficiently low,  the particles involved in neutrino mass generation could have an impact 
in neutrino oscillations. 
Atmospheric neutrinos provide an ideal tool to test new physics, as their spectrum covers a huge energy range ($\sim 1 - 10^5$ GeV) and they may travel distances across the Earth from tens to several thousands 
kilometers, for different zenith angles. 
While most previous analyses have focused on relatively low-energy, $\cal{O}$(10 GeV), atmospheric 
neutrino data, in the following we discuss the potential of the  
high energy ($>$ 100 GeV) atmospheric neutrinos at IceCube to 
set constraints on light  sterile neutrinos  and 
non-standard interactions (NSI) in the $\nu_\mu - \nu_\tau$ sector.

\section{Sterile neutrino searches}

Sterile neutrinos with mass in the eV range can account for the anomalies found in
short-baseline accelerator (LSND, MiniBooNE), reactor and gallium (with high intensity radioactive sources ) oscillation experiments \footnote{See talks by A. Palazzo and C. Buck in this conference for details about 
such anomalies.}.

Atmospheric neutrino data would be affected by additional (beyond 3 flavour) oscillations into sterile neutrinos
in this mass range.
At high energies, $E_\nu > 100$ GeV, oscillations due to the known atmospheric and solar mass splittings 
have wavelengths larger than the Earth diameter and can be neglected, while matter effects can enhance 
the transition between active and sterile neutrinos  with $\Delta m^2 = 0.01$ eV$^2$ - 10 eV$^2$, leading to 
detectable  energy and zenith angle distortions  of the neutrino flux.

\begin{figure}[htb]
\centering
\includegraphics[height=2.5 in]{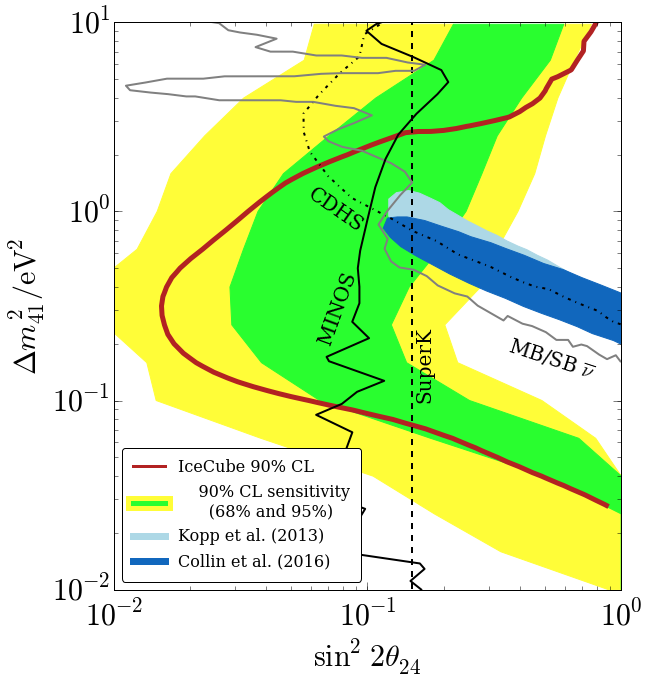}
\caption{Results from the IceCube search for sterile neutrinos.
The orange solid line is the 90\% CL contour, while the bands contain 68\% (green) and 95\% (yellow) of the 90\% countours in simulated pseudo-experiments. 90\% CL exclusion limits from previous experiments
are also shown, as well as 99\% CL allowed regions from global fits to appearance experiments, including 
MiniBooNE and LSND. Figure taken from \cite{TheIceCube:2016oqi}.}
\label{fig:sterile}
\end{figure}

 The IceCube collaboration has performed a search for $(\nu_\mu + \bar{\nu}_\mu)$ disappearance
 through oscillation into a sterile neutrino  using the publicly available IceCube one-year upgoing muon 
 sample IC86 (IceCube 86-string configuration), which contains 20145 muons corresponding to 
 atmospheric neutrinos in the  approximate energy range 320 GeV to 20 TeV \cite{TheIceCube:2016oqi}.
 The full active and sterile 
 neutrino evolution has been performed    by employing the $\nu$-SQuIDS package.
The resulting 90\% CL exclusion limits are shown in Fig.~\ref{fig:sterile}, together with  limits from previous experiments and global fits for reference.

\section{Neutrino non-standard interactions}

For recent reviews about NSI, and a complete list of references,  see, e.g., Refs.~\cite{Ohlsson:2012kf, Miranda:2015dra}.

\subsection{Theory}

Neutral current (NC) neutrino  NSI, also called matter NSI, can be  parametrized via model-independent, effective
four-fermion operators as follows:
\beq
\label{eq:NC}
{\cal L}^{\rm NC}_{\rm NSI} = - 2 \sqrt{2} \, G_F \, \varepsilon^{fP}_{\alpha \beta} \, 
 (\bar{\nu}_\alpha \gamma_{\rho} L \nu_\beta) (\bar{f} \gamma^{\rho}P  f) ~,
\eeqn
where $\varepsilon^{fP}_{\alpha \beta}$ are the NC NSI parameters (by hermiticity $\varepsilon^{fP}_{\alpha \beta} = (\varepsilon^{fP}_{\beta \alpha})^*$), $P=\{L,R\}$ (with $L$ and $R$ the left and right quirality projectors) and $f$ is any SM fermion;
charged-current (CC) NSI can be described analogously.
Model-independent bounds on CC NSI, which affect neutrino's production and detection, are generally one order of magnitude stronger than NC ones  \cite{Biggio:2009nt}, that mainly modify neutrino propagation;
 thus we neglect CC NSI in the following.

It is desirable that the four-fermion vertices in eq.~(\ref{eq:NC}) arise
in an $SU(2) \times U(1)_Y$ gauge invariant theory, where they can be generated by
operators of dimension six, eight and larger.
In general, new physics which induces the dimension 6 operator also
induces an operator involving charged leptons, with a coefficient of
the same order by $SU(2)$
invariance. Charged lepton physics imposes
tight constraints on these coefficients of dimension 6 operators, 
rendering neutrino NSI unobservable. 
There are only two UV completions (at tree level) in which neutrino NC NSI can be induced 
by dimension six operators without the charged-lepton counterpart, and without fine-tuned ad-hoc cancellations:
one $SU(2)$ singlet scalar with $Y=1$ \cite{Bilenky:1993bt} and non-canonical neutrino kinetic terms due to 
mixing with heavy SM singlet fermions which are integrated out \cite{Antusch:2008tz}. 
In this last case, after diagonalising and  normalising the neutrino kinetic terms, a non-unitary lepton mixing matrix is generated  that leads to 
NC NSI just for neutrinos. However, a detailed study of this class of  scenarios shows that the constraints on the NC NSI  turn out to be even stronger than the ones for operators which also produce interactions of four charged fermions at the same level: typically  
$\varepsilon^{fP}_{\alpha \beta}  < {\cal O}(10^{-3})$, too small to be observable in current 
neutrino oscillation experiments. 
The only exception is the case of  non-unitarity effects produced by mixing with sterile neutrinos 
in the keV range \cite{Blennow:2016jkn}, which allows for NC NSI parameters of ${\cal O}(10^{-2})$ (see also talk by J. Lopez-Pavon in this conference).

 At dimension 8 or larger, in principle an operator as in equation~(\ref{eq:NC}) can appear at tree
level without any charged lepton counterpart and generate sizeable NC NSI;
in practice, constructing ($SU(2)_L \times U(1)_Y$ gauge-invariant) UV completions  with large neutrino NSI and consistent with all current experimental constraints, requires a certain amount of fine-tuning
\cite{Gavela:2008ra}, although they cannot be completely excluded.

Recently, it has been considered the possibility of generating the NC NSI in models based on  a new $U(1)'$ 
gauge interaction with a light gauge boson mass $\sim $ 10 MeV. Since for neutrino propagation only forward scattering is relevant, the effective coupling in eq.~(\ref{eq:NC}) can be used even for neutrino energies 
much higher than the mediator mass, while in scattering experiments such as NuTeV the effects are 
stronlgy suppressed, allowing to satisfy current bounds while having potentially sizeable NC NSI
\cite{Farzan:2015doa}.

\subsection{Phenomenology}

 In the presence of NC NSI, the effective Hamiltonian that controls neutrino propagation in matter can be written as 
 \beq
\label{eq:hnsi}
H(E_\nu,x) = \frac{1}{2 E_\nu} U M^2 U^\dagger + {\rm diag}(V_e,0,0)  + \sum_f V_f  \, \varepsilon^{fV}  \ , 
\eeqn
where $U$ is the PMNS mixing matrix, $M^2 = {\rm diag}(0, \Delta m_{21}^2, \Delta m_{31}^2)$,  with  $\Delta m_{ij}^2 \equiv m_i^2 - m_j^2$ the neutrino mass square differences and $V_f (x)=  \sqrt{2} \, G_F\,  n_f(x)$, with $n_f(x)$ the number density of fermion $f$.
The effect of NSI is encoded in the last term of Eq.~(\ref{eq:hnsi}), where $\varepsilon^{fV}$ is the matrix in lepton flavor space that contains the vector combination of the NSI chiral parameters, $\varepsilon^{fV}_{\alpha \beta}  =  \varepsilon^{fR}_{\alpha \beta}  +  \varepsilon^{fL}_{\alpha \beta}$. 
For antineutrinos, the  matter potentials change sign, $V_f \to - V_f$, and $U \to U^*$. It is convenient to define effective NSI parameters for a given medium by normalizing the fermion number density, $n_f$, to the density of $d$-quarks, $n_d$,
\beq 
\varepsilon_{\alpha \beta}  \equiv \sum_{f}  \frac{n_f}{n_d} \, \varepsilon^{fV}_{\alpha \beta} ~, 
\eeqn
so that $\sum_f V_f  \varepsilon^{fV} \equiv  V_e \, r \, \varepsilon = V_d \, \varepsilon$, and $r = n_d/n_e$. For the Earth, $n_n \approx n_p$ and therefore, $r \approx 3$.
Notice that oscillation experiments are only sensitive to the differences between the diagonal terms in the matter potential, e.g., 
$ \varepsilon'_{\alpha\alpha} \equiv \varepsilon_{\alpha\alpha} - \varepsilon_{\mu\mu}$.
 
The upper bounds on $\varepsilon^{fP}_{\alpha\beta}$ from neutrino oscillation and scattering data  are 
rather weak.
Even more, in addition to the standard LMA solution to solar neutrino data, there
is another solution called LMA-Dark which requires NSI with effective couplings 
$\varepsilon^{qV}_{ee} - \varepsilon^{qV}_{\mu\mu}$ as large as the SM ones, 
as well as a solar mixing angle in the second octant, and implies an ambiguity in the neutrino mass ordering 
\cite{Miranda:2004nb,Gonzalez-Garcia:2013usa}. 
The degeneracy between the two solutions can be lifted by 
a combined analysis of data from oscillation experiments with the neutrino scattering experiments CHARM and NuTeV, provided the neutrino NSI take place with down quarks, 
and  the mediators are not much lighter than the electroweak scale \cite{Coloma:2017egw}.
For light mediators, the LMA-Dark solution can be ruled out  at DUNE \cite{Coloma:2015kiu} or in future coherent neutrino-nucleus scattering experiments \cite{Coloma:2017egw}.

Off-diagonal NSI $\varepsilon^{qV}_{e\tau} \sim {\cal O}(0.1)$ is also slightly favoured, due to  
the suppression of the upturn on low energy spectrum of solar neutrinos, which is in a mild tension with the standard neutrino oscillation scenario \cite{Gonzalez-Garcia:2013usa}; such NSI can be tested by 
atmospheric neutrinos at Hyper-Kamiokande \cite{Fukasawa:2016nwn}.

Many atmospheric neutrino's NSI analysis restrict to the $\nu_\mu - \nu_\tau$ sector, however allowing for 
all non-vanishing $\varepsilon_{\alpha\beta}$ in the 
$\nu_e - \nu_\tau$ sector leads to a matter potential that mimics vacuum oscillations 
$\nu_\mu \rightarrow \nu_{\tau'}$ with the same $E_\nu$
dependence, but modified mixing and mass differences, along the parabola $\varepsilon_{\tau\tau} = |\varepsilon_{e\tau}|^2/(1 + \varepsilon_{ee})$.
As a consequence, ${\cal O}(1)$ values of $\varepsilon_{\tau\tau}, \varepsilon_{e\tau}$ are possible 
in this region. We disregard this somehow fine-tuned possibility and consider 
the effect of $\nu_\mu - \nu_\tau$  NSI in the high energy atmospheric neutrino sample at IceCube.

\subsection{NSI with atmospheric neutrinos at IceCube}

This section is based on  \cite{Salvado:2016uqu}, where details of the analysis can be found.
The relative size of NSI with respect to standard neutrino oscillations depends on the neutrino energy, 
therefore atmospheric neutrino data provides the possibility of
exploiting the NSI energy dependence over a large range of energies and baselines 
in order set stronger constraints.

The standard evolution Hamiltonian for neutrinos in a medium includes the coherent forward scattering on fermions of the type $f$, $\nu_\alpha + f \to \nu_\beta + f$, given by the matter interaction potential in Eq.~(\ref{eq:hnsi}), 
which affects neutrino oscillations. 
As the neutrino-nucleon cross section increases with energy, for energies above $\sim$TeV, the neutrino flux gets attenuated: neutrinos are absorbed via CC interactions and redistributed (degraded in energy)  via NC 
ones. On the other hand, the  $\nu_\tau$ flux regeneration effect is negligible
for  the high-energy IceCube sample of atmospheric neutrinos analysed.

\begin{figure}
	\centering
	\includegraphics[width=0.49\textwidth]{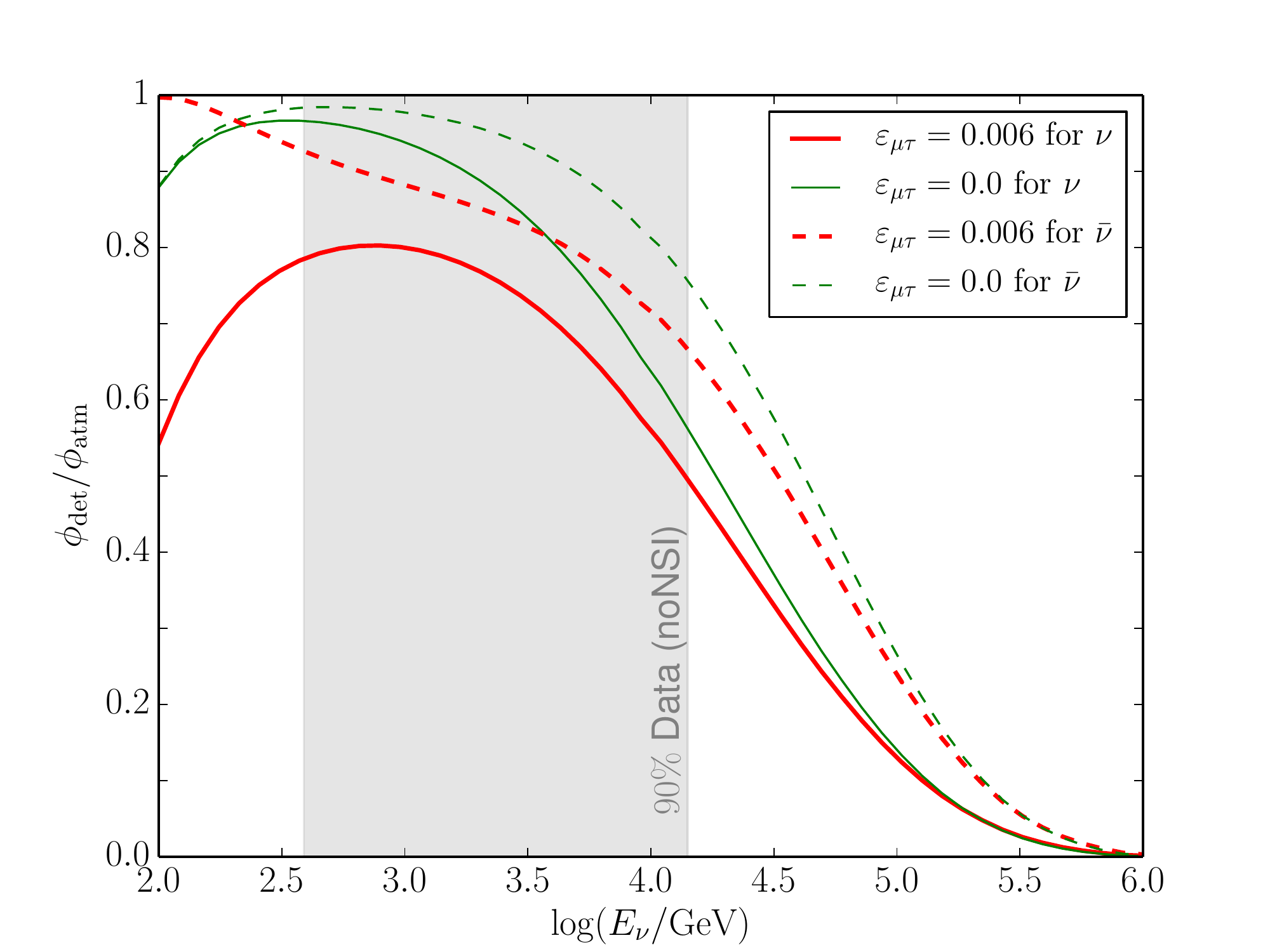}  
	\includegraphics[width=0.49\textwidth]{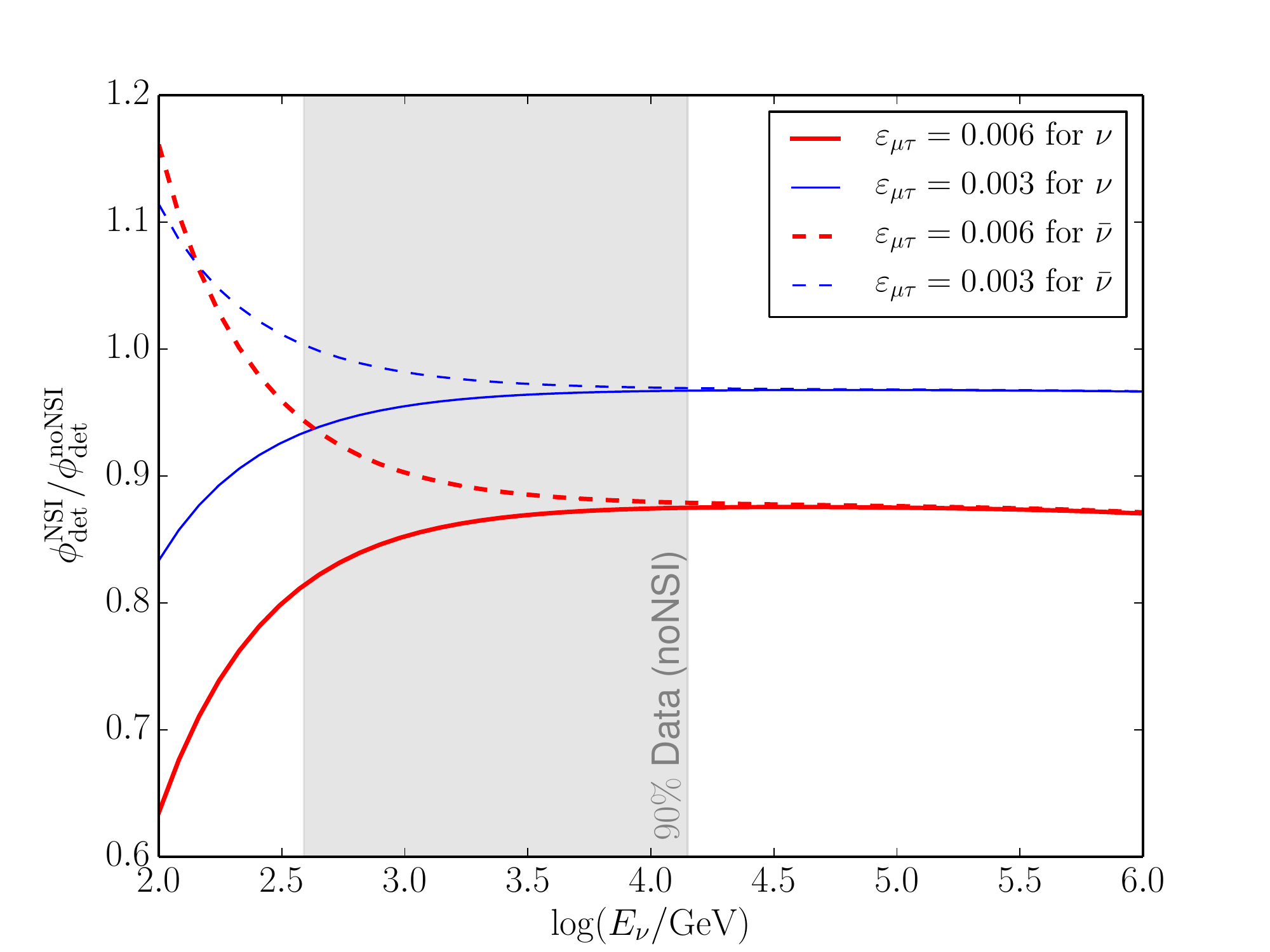}  
	\caption{{\it Left panel:} Comparison of the ratios of propagated to unpropagated atmospheric $\nu_\mu$ (solid lines) and $\bar \nu_\mu$ (dashed lines) fluxes for  $\varepsilon_{\mu \tau} = 0.006$ (thick red lines) and $\varepsilon_{\mu\tau} = 0$ (thin green lines). {\it Right panel:} Comparison of the ratios of atmospheric 
	$\nu_\mu$  and $\bar \nu_\mu$ fluxes at the detector (after propagation) with NSI to those without NSI, for two values of $\varepsilon_{\mu\tau}$. In both panels, $\cos \theta_z = -1$ and $\varepsilon' = 0$. The gray area corresponds to the energy interval that produced 90\% of the events in the entire sample considered here in the absence of NSI effects.}
	\label{fig:fratio_1D}
\end{figure}

We have used the density matrix formalism to describe the neutrino propagation 
though the Earth, including SM NC and CC inelastic scattering. We have solved numerically the full three-neutrino evolution  equation by employing the publicly available libraries SQuIDS and $\nu$-SQuIDS. 
In Fig.~\ref{fig:fratio_1D} we show the effect of attenuation and NSI for both, neutrinos and antineutrinos.
Notice that al low energies, the effect of NSI and attenuation is different for neutrinos and antineutrinos, while 
at high energies both ratios coincide (right panel).

In order to understand this behaviour, it is illustrative to study analytically the oscillation probabilities for two 
neutrinos  in the
approximation of constant matter density and neglecting inelastic scattering. 
When vacuum and matter NSI terms are of the same order of magnitude
($\Delta m_{31}^2/2 E_\nu \sim V_{\rm NSI} $, with $V_{\rm NSI} =V_d \sqrt{4\epsilon_{\mu\tau}^2 + \epsilon'^2}$), 
the transition probability after propagating a distance $L$ reads
\beq
P(\nu_\mu \to \nu_\tau) \simeq \left(\sin 2\theta_{23} \, \frac{\Delta m^2_{31}}{2 \, E_\nu} +  2 \, V_d \, \varepsilon_{\mu \tau}\right)^2 \left(\frac{L}{2}\right)^2  ~,
\eeqn
while the NSI matter term has opposite sign for antineutrinos. 
However in the high-energy limit the matter NSI term dominates over vacuum oscillations, and for 
$V_{\rm NSI}/L \ll 1$ the two-neutrino transition probability is approximately given by
\beq 
P(\nu_\mu \to \nu_\tau) \simeq \left(\sin^2 2\xi\right) \, \phi_{\rm mat}^2 =  (\varepsilon_{\mu \tau} \, V_d \, L)^2 
\ , 
\eeqn
which is proportional to $\varepsilon_{\mu \tau}^2$ and becomes independent of $\epsilon'$. 
As a consequence, the high-energy IceCube atmospheric neutrino data cannot significantly 
constrain the diagonal NSI parameter $\epsilon'$, so in our analysis we use a prior on $\epsilon'$ 
based on SK limits \cite{Mitsuka:2011ty}, which were obtained from data at lower energies:
$|\varepsilon'| = | \varepsilon_{\tau \tau} - \varepsilon_{\mu \mu}| < 0.049$  at 90\% confidence level (C.L.).
From these results, we 
set the $1\sigma$ C.L. prior on $\varepsilon'$ to $\sigma_{\varepsilon'} = 0.040$.

In our analysis   we use 
the same IceCube data sample as in the search for light sterile neutrino signatures described in 
sec. 2 ~\cite{TheIceCube:2016oqi}.
 In order to perform the analysis, we used the public IceCube Monte Carlo\footnote{{https://icecube.wisc.edu/science/data/IC86-sterile-neutrino}} that models the detector realistically and allows us to relate physical quantities, as the neutrino energy and direction, to observables, as the reconstructed muon energy and zenith angle. 

To evaluate the impact of possible systematic uncertainties, we have included the following nuisance parameters:  
normalization of the atmospheric neutrino flux, $N$, pion-to-kaon ratio in the atmospheric neutrino flux, $\pi/K$,  spectral index of the atmospheric neutrino spectrum, $\Delta \gamma$, uncertainties in the efficiency of the digital optical modules of the detector, DOM$_{\rm eff}$
and current uncertainties in $\Delta m^2_{31}$ and $\theta_{23}$. In addition, 
we have considered several combinations of primary cosmic-ray flux and hadronic interaction models, 
being our default choice the  Honda-Gaisser model and Gaisser-Hillas H3a correction (HG-GH-H3a) for the primary cosmic-ray flux and the QGSJET-II-4 hadronic model.
We show in the right panel of Fig.~\ref{fig:systematics} the effect of this source of uncertainty in the posterior probabilities of
$ \varepsilon_{\mu \tau} $.

In this way we have obtained the most up-to-date limits on the off-diagonal NSI parameter $\varepsilon_{\mu \tau}$ and showed they currently depend very little on the systematic uncertainties. For our default 
combination of models, 
we find 
\beq
\label{eq:bound}
- 6.0 \times 10^{-3} < \varepsilon_{\mu \tau} < 5.4 \times 10^{-3} ~, \hspace{1cm} \textrm{90\% credible interval (C.I.)},
\eeqn
and similar results from all the other possible combinations. Our bound  is  comparable to the one obtained in
\cite{Esmaili:2013fva}, using 79-string IceCube configuration and DeepCore data, although they do not 
include nuisance parameters in their analysis.

\section{Summary}

\begin{figure}
	\centering
	\includegraphics[width=0.49\textwidth]{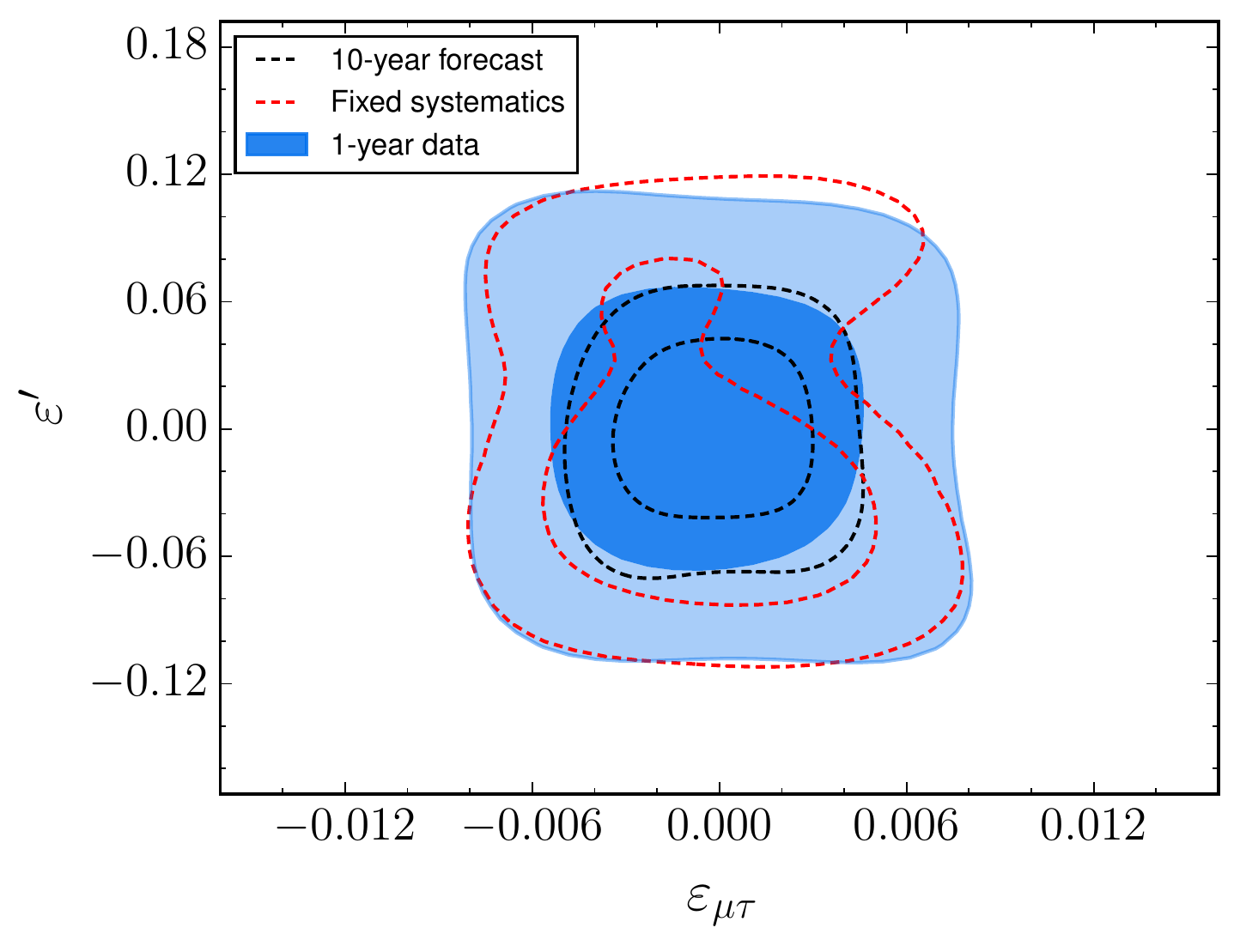}
	\includegraphics[width=0.46\textwidth]{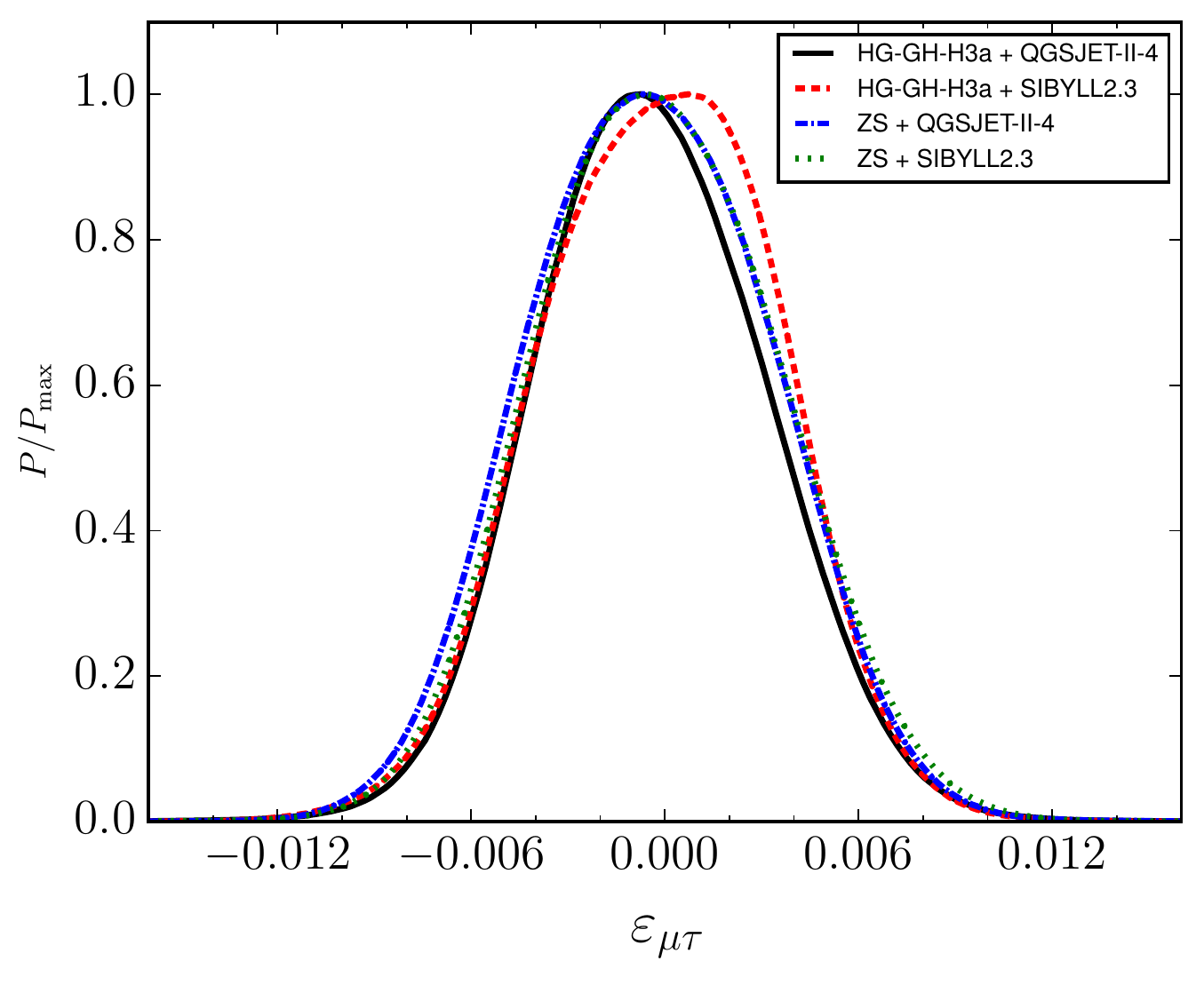}
	\caption{{\it Left panel:} Comparison of the 68\% and 95\% credible contours in the $\varepsilon_{\mu \tau}-\varepsilon'$ plane for our default analysis (filled blue regions) with those obtained when all nuisance parameters are fixed at their default values (red closed curves). We also show the result expected in the case of no NSI after 10 years of data taking (black closed curves). {\it Right panel:} Posterior probabilities of $\varepsilon_{\mu \tau}$, after marginalizing with respect to the rest of parameters, for the four combinations of primary cosmic-ray spectrum and hadronic models.}
\label{fig:systematics}
\end{figure}

We have described two recent examples of the potential of high-energy atmospheric neutrino data  from IceCube
to constrain new physics,
namely the search for sterile neutrinos by the IceCube collaboration  (Fig.~\ref{fig:sterile}) and
 the limits on off-diagonal $\nu_\mu - \nu_\tau$  NSI, both using the one-year upgoing muon sample, IceCube 86-string configuration.
We have obtained the limit 
$- 6.0 \times 10^{-3} < \varepsilon_{\mu \tau} < 5.4 \times 10^{-3}$ (90\% credible interval),
and showed that  
systematics currently affect very little this bound. 
We also provide a forecast of the future sensitivity to NSI by simulating 10~years of high-energy neutrino data in IceCube. 
Fig.~\ref{fig:systematics} summarizes our findings.

\Acknowledgements
I would like to  thank the organizers for the interesting conference, and my NSI collaborators, Olga Mena, Sergio Palomares-Ruiz and Jordi Salvado. 
I am also grateful to Marian T\'ortola and Yasaman Farzan for useful discussions about some of the topics
presented in this talk.

\end{document}